\documentclass[fleqn]{article}
\usepackage{amsmath,amssymb}
\usepackage{epsfig,graphicx}

\catcode`@=11 \@addtoreset{equation}{section} \catcode`@=12

\begin{document}
\title{\vskip-1.7cm \bf  New representation of the nonlocal ghost-free gravity theory}
\date{}
\author{A.O.Barvinsky$^{1}$ and Yu.V.Gusev$^{2,\,3}$}
\maketitle \hspace{-8mm} {\,\,$^{1}$\em
Theory Department, Lebedev
Physics Institute, Leninsky Prospect 53, Moscow 119991, Russia\\
$^{2}$Lebedev Research Center in Physics, Leninsky Prospect 53, Moscow 119991, Russia\\
$^{3}$IRMACS Centre, Simon Fraser University,
8888 University Drive, Burnaby, B.C. V5A 1S6, Canada}

\begin{abstract}
A new representation is found for the action of the recently suggested ghost-free nonlocal gravity models generating de Sitter or Anti-de Sitter background with an arbitrary value of the effective cosmological constant. This representation allows one to extend applications of these models from maximally symmetric to generic Einstein spaces and black hole solutions, but clearly indicates violation of the general relativistic limit in this class of theories, induced by their infrared behavior. It is shown that this limit can be recovered in a special conformal frame of these theories, and their relation to critical gravity models is also briefly discussed.
\end{abstract}

\maketitle

\section{Introduction}
A new approach to the dark energy problem, that has recently been suggested in \cite{serendipity}, is inspired by the necessity to avoid the fine tuning problem. This approach suggests the theory in which the de Sitter or anti-de Sitter evolution can occur at any value of the effective cosmological constant $\Lambda$ -- the antithesis to the dark energy scale encoded in the action of the model and fine tuned to the observational data. A concrete observable value of $\Lambda$ in this theory is supposed to be selected by the mechanism analogous to symmetry breaking \cite{serendipity}. Interestingly, the realization of this approach quite unexpectedly has also led to the analogue of the dark matter phenomenon characterized at large distances by gravitational attraction stronger than in general relativity or Newton theory.

The action of this theory was shown to generate vacuum equations of motion which have as a solution the de Sitter or anti-de Sitter background. This background carries only transverse-traceless gravitons as propagating physical modes and is free from ghost instabilities. The stability property was proven in \cite{serendipity} by very extensive calculations for a maximally symmetric background and then extended in \cite{Solodukhin} to generic Einstein spaces $R_{\mu\nu}=\Lambda g_{\mu\nu}$ with a vanishing traceless part of the Ricci tensor
    \begin{eqnarray}
    E_{\mu\nu}\equiv
    R_{\mu\nu}-\frac14\,g_{\mu\nu}R=0.  \label{Einsteinspace}
    \end{eqnarray}
Thus, this model could be regarded as one of the first modifications of the Einstein theory made by Einstein himself, who for reasons of unification with electromagnetism suggested to replace the Einstein tensor $G_{\mu\nu}=R_{\mu\nu}-\frac12g_{\mu\nu}R$ in the left hand side of Einstein equations by $E_{\mu\nu}$ \cite{Einstein}.

The action with these properties is the following nonlocal functional of the spacetime metric $g_{\mu\nu}$,\footnote{We use the Euclidean signature spacetime and curvature tensor conventions, $R=g^{\mu\nu}R_{\mu\nu}=g^{\mu\nu}R^\alpha_{\;\;\mu\alpha\nu}= g^{\mu\nu}\partial_\alpha\Gamma^\alpha_{\nu\mu}
-...\;$.}
    \begin{eqnarray}
    &&S=\frac{M^2}2\int dx\,g^{1/2}\,\left\{-R+
    \alpha\,R^{\mu\nu}
    \frac1{\Box+\hat P}\,G_{\mu\nu}
    \right\},\;\;\;\;                        \label{action}\\
    &&\hat P\equiv P_{\alpha\beta}^{\;\;\;\mu\nu}
    =a R_{(\alpha\;\;\beta)}^{\;\;\,(\mu\;\;\,\nu)}
    +b \big(g_{\alpha\beta}R^{\mu\nu}
    +g^{\mu\nu}R_{\alpha\beta}\big)
    +c R^{(\mu}_{(\alpha}\delta^{\nu)}_{\beta)}
    +d R\,g_{\alpha\beta}g^{\mu\nu}
    +e R \delta^{\mu\nu}_{\alpha\beta},     \label{potential}
    \end{eqnarray}
where the hat denotes matrices acting on symmetric tensors, and we use the condensed notation for the Green's function of the covariant operator
    \begin{eqnarray}
    \Box+\hat P\equiv\Box\,\delta_{\alpha\beta}^{\;\;\;\mu\nu}
    +P_{\alpha\beta}^{\;\;\;\mu\nu},
    \quad
    \Box=g^{\lambda\sigma}
    \nabla_\lambda\nabla_\sigma,  \label{operator}
    \end{eqnarray}
acting on any symmetric tensor field $\Phi_{\mu\nu}$ as
    \begin{eqnarray}
    &&\frac1{\Box+\hat P}\,\Phi_{\mu\nu}(x)\equiv
    \Big[\,\frac1{\Box+\hat P}\,\Big]_{\mu\nu}^{\alpha\beta}\Phi_{\alpha\beta}(x)
    =\int dy\,G_{\mu\nu}^{\alpha\beta}(x,y)\,
    \Phi_{\alpha\beta}(y)
    \end{eqnarray}
with $G_{\mu\nu}^{\alpha\beta}(x,y)$ -- the two-point kernel of this Green's function.

Thus this model formally falls into the category of nonlocal theories descending from the old approach to nonlocal QFT and quantum gravity \cite{Efimov} and its latest development \cite{nonloccosm} motivated by cosmological implications \cite{DeffWood,TsamisWoodard,Odintsovetal,ParkDodelson} and the requirements of renormalizability and unitarity \cite{latest}. However, in contrast to the functional ambiguity in the choice of action, characteristic of these works, here we have only a parametric freedom. The action (\ref{action}) has one dimensional parameter $M$ and six dimensionless parameters $\alpha$, $a$, $b$, $c$, $d$ and $e$, the first one $\alpha$ determining the overall magnitude of the nonlocal correction to the Einstein term. For a small value of $|\alpha|\ll 1$ and the value of $M$ related to the Planck mass $M_P$,
    \begin{eqnarray}
    M^2=\frac{M^2_P}{1-\alpha},  \label{M_Prenorm}
    \end{eqnarray}
the theory (\ref{action}) has a GR limit on a {\em flat-space background}\footnote{Note that the nonlocal part contributes to the quadratic part of the action in metric perturbations and renormalizes the value of the Newton constant \cite{covnonloc}. The structure of nonlocal corrections in (\ref{action}) is motivated by the nonlocal covariant expansion in powers of the curvature for the Einstein action including the Gibbons-Hawking surface term \cite{covnonloc}.}, whereas the rest of the parameters are restricted by the requirement of a stable (A)dS solution existing in this theory. These restrictions read
    \begin{eqnarray}
    &&\alpha=-A-4B,    \label{relation}\\
    &&C=\frac23,        \label{Crelation}\\
    &&M_{\rm eff}^2
    =\frac{A^2-\alpha^2}{\alpha}\,M^2>0. \label{effectivemass}
    \end{eqnarray}
where the new quantities $A$, $B$ and $C$ equal in terms of original parameters
    \begin{eqnarray}
    &&A=a+4\,b+c,\quad
    B=b+4\,d+e,                     \label{AB}\\
    &&C=\frac{a}3-c-4e,            \label{C}
    \end{eqnarray}
and $M_{\rm eff}$ is the effective Planck mass which determines the cutoff scale of perturbation theory in the (A)dS phase and the strength of the gravitational interaction of matter sources.

The condition (\ref{relation}) guarantees the existence of the (A)dS solution, while Eqs.(\ref{Crelation})-(\ref{effectivemass}) are responsible for its stability.  The calculation of the gauge fixed quadratic part of the action on the (A)dS background shows that longitudinal and trace modes which formally have a ghost nature are unphysical and can be eliminated by residual gauge transformations preserving the gauge \cite{serendipity}. This well-known mechanism leaves only two transverse-traceless physical modes propagating on the (A)dS background, similar to GR theory. Finally, as was shown in \cite{Solodukhin} the additional condition,
    \begin{eqnarray}
    a=2,                                 \label{arelation} \end{eqnarray}
allows one to extend the ghost stability arguments to generic Einstein backgrounds with a nonvanishing Weyl tensor.

What is critically different from the GR phase of the theory -- its effective gravitational constant $G_{\rm eff}\equiv 1/8\pi M_{\rm eff}^2$ which can be much larger than than the Newton constant $G_N=1/8\pi M_P^2$, because in view of (\ref{relation}) a natural range of the parameter $A$ is $A\sim\alpha$, and $G_{\rm eff}\sim G_N/|\alpha|\gg G_N$. This property can be interpreted as a simulation of DM mechanism, because it implies strengthening of the gravitational attraction in the (A)dS phase of the theory and its possible effect on rotation curves at relevant distance scales.

The main goal of this paper is a simple derivation of the above results, which is based on the new representation of the action (\ref{action}) with a critical value (\ref{relation}) of $\alpha$
    \begin{eqnarray}
    &&S=-\frac{M^2_{\rm eff}}2
    \int dx\,g^{1/2}\,E^{\mu\nu}
    \frac1{\Box+\hat P}\,E_{\mu\nu}.  \label{newrep}
    \end{eqnarray}
As we show below, it holds for closed compact spacetimes with the Euclidean signature. This Euclidean setting underlies the problems of black hole thermodynamics and the Schwinger-Keldysh technique for quantum expectation values in a special class of quantum states like Euclidean (quasi-de Sitter invariant) vacuum. The advantage of this representation is obvious -- quadratic in $E_{\mu\nu}$ form of (\ref{newrep}) directly indicates the existence of Einstein space solutions satisfying (\ref{Einsteinspace}) and also very easily gives the inverse propagator of the theory on their background. Single-pole nature of the propagator with a positive residue yields the ghost-free criteria (\ref{Crelation})-(\ref{effectivemass}) and (\ref{arelation}). All this is presented in the next two sections.

The concluding section is devoted to the discussion of a serious difficulty of our model, which clearly manifests itself in its new representation (\ref{newrep}). In contrast to anticipations of \cite{serendipity}, the theory has the GR limit neither in the short wavelengths regime $\nabla\nabla\gg R$ nor in the limit of $\alpha\to 0$. This property, that was first observed in \cite{Solodukhin}, is explained by the presence of the constant zero mode of the scalar $\Box$ operator on a compact spacetime without a boundary. This leads to a nonanalytic behavior of the theory at $\alpha\to 0$ and the absence of a crossover between its dark energy phase and the GR phase, the latter existing only in the asymptotically flat spacetime. Then we discuss the possibility to recover the GR phase in a special conformal frame of the spacetime metric.

Though a direct application of the model (\ref{action}) in realistic cosmology seems questionable, it might be interesting in context of currently popular critical gravity theories \cite{critical}. In particular, it looks like a nonlocal version of these theories quadratic in curvature, because for a critical value of $\alpha$ (\ref{relation}) and a generic constant $C\neq 2/3$, its propagator has a double pole nature and incorporates the so-called logarithmic modes \cite{critical}.

\section{Euclidean field theory vs Schwinger-Keldysh technique and compactness of spacetime}

The action (\ref{action}) above requires specification of boundary conditions for the Green's function.  Any choice, however, violates causality in the initial value problem for a dynamically evolving fields. Their nonlocal equations of motion break causality because the behavior of the field at any spacetime point is not independent of the field values in the future light cone of this point \cite{serendipity}. Therefore, applicability of this action is restricted to the class of problems alternative to those of the evolution from the initial state. One such class is represented by gravitational thermodynamics implemented by the Euclidean quantum gravity (EQG) -- quantum gravity in the Euclidean signature spacetime.

Another class in the Lorentzian signature spacetime is mediated by a special technique adapting nonlocal equations of motion to causality. This is the Schwinger-Keldysh technique \cite{SchwKeld} for quantum expectation values $\langle\,{IN}\,|\,\hat{\cal O}(x)\,|\,{ IN}\,\rangle$ of local physical observables $\hat{\cal O}$ in the initial quantum state $|\,{IN}\,\rangle$. Though the equations for $\langle\,{IN}\,|\,\hat{\cal O}(x)\,|\,{IN}\,\rangle$ are nonlocal, this quantity depends only on the past of the point $x$. This property is again not manifest and turns out to be the consequence of locality and unitarity of the original fundamental field theory (achieved via a complicated set of cancelations between nonlocal terms with chronological and anti-chronological boundary conditions). In contrast to Wick rotation in the S-matrix theory this technique is not related to Euclidean quantum field theory and to EQG, in particular.

However, there exists a class of problems for which a retarded nature of effective equations explicitly follows from their quantum effective action calculated in Euclidean spacetime \cite{beyond}. This is a statement based on Schwinger-Keldysh technique \cite{SchwKeld,SchwKeld2} that for an appropriately defined initial quantum state $|{\rm in}\rangle$ the effective equations for the mean field
    \begin{eqnarray}
    g_{\mu\nu}=\langle{\,IN}\,|\,\hat g_{\mu\nu}|\,{IN}\,\rangle
    \end{eqnarray}
originate from the {\em Euclidean} quantum effective action $S=S_{\rm Euclidean}[g_{\mu\nu}]$ by the following procedure \cite{beyond}\footnote{We formulate this statement directly for the case of gravity theory with the expectation value of the metric field operator $\hat g_{\mu\nu}(x)$, though it is valid in a much wider context of a generic local field theory \cite{beyond}.}. Calculate nonlocal $S_{\rm Euclidean}[g_{\mu\nu}]$ and its variational derivative. In the Euclidean signature spacetime nonlocal quantities, relevant Green's functions and their variations are generally uniquely determined by their trivial (zero) boundary conditions at infinity, so that this variational derivative is unambiguous in Euclidean theory. Then make a transition to the Lorentzian signature and impose the {\em retarded} boundary conditions on the resulting nonlocal operators,
    \begin{eqnarray}
    \left.\frac{\delta S_{\rm Euclidean}}{\delta g_{\mu\nu}(x)}\right|_{\;++++\,\;
    \Rightarrow\;-+++}^{\;\rm retarded}=0.   \label{EuclidLorentz}
    \end{eqnarray}
These equations are causal ($g_{\mu\nu}(x)$ depending only on the field behavior in the past of the point $x$) and satisfy all local gauge and diffeomorphism symmetries encoded in the original $S_{\rm Euclidean}[g_{\mu\nu}]$.

To be more precise, the relation (\ref{EuclidLorentz}), that was proven to the first order of perturbation theory in \cite{Hartle-Horowitz} and to all orders in \cite{beyond}, originates as follows. The one-loop equation for the mean IN-IN field $g(x)$ contains the tadpole type quantum contribution
    \begin{eqnarray}
    &&\frac{\delta S}{\delta g(x)}
    +\frac{i}2\,\int dy\,dz\,\frac{\delta^3 S}{\delta g(x)\,\delta g(y)\,\delta g(z)}\,G_{IN\!-\!IN}(y,z)=0,\\
    &&\nonumber\\
    &&G_{IN\!-\!IN}(x,y)=
    \langle IN\,|\,\hat g(x)\,\hat g(y)\,|\,IN\rangle,
    \end{eqnarray}
with the IN-IN Wightman Green's function $G_{IN\!-\!IN}(x,y)$ alternative to the conventional Feynman propagator. As was shown in \cite{beyond} for the Poincare invariant vacuum state (associated with a plane wave decomposition of the IN-operators) the following relation holds
    \begin{eqnarray}
    &&\frac{i}2\,\int dy\,dz\,\frac{\delta^3 S}{\delta g(x)\,\delta g(y)\,\delta g(z)}\,G_{IN\!-\!IN}(y,z)=
    \left.\frac{\delta\varGamma_{E}^{\rm 1-loop}}
    {\delta g(x)}\,\right|_{\;++++\,\Rightarrow\,-+++}^{\;\rm retarded}\\
    &&\varGamma_{E}^{\rm 1-loop}=\frac12\,{\rm Tr}\ln\frac{\delta^2 S_{\rm Euclidean}}{\delta g(x)\,\delta g(y)}.
    \end{eqnarray}
This confirms the relation (\ref{EuclidLorentz}) with the full one-loop Euclidean effective action $\varGamma_{\rm Euclidean}=S_{\rm Euclidean}+\varGamma_{E}^{\rm 1-loop}$.

\begin{figure}[h]
\centerline{\epsfxsize 12cm \epsfbox{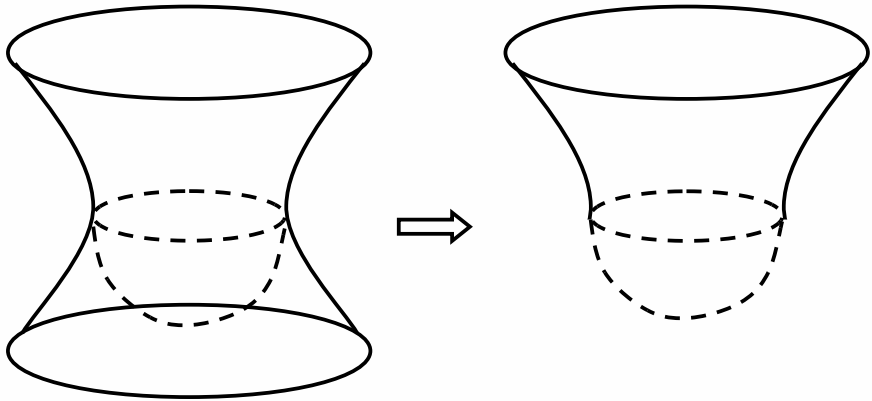}} \caption{\small
Euclidean de Sitter hemisphere denoted by dashed lines is used as a tool for constructing the Euclidean de Sitter invariant vacuum by the path integral over regular fields on the resulting compact spacetime.
 \label{Fig.1}}
\end{figure}

In \cite{serendipity} it was assumed that the model with the action (\ref{action}) falls into the range of validity of this procedure, and the action itself coincides with the nonlocal effective action of the Euclidean QFT calculated within certain approximation of the curvature expansion \cite{quantum0,quantum1}. This assumption implies a particular vacuum state $|{IN}\rangle$ and the one-loop approximation (in which it was proven to the first order of perturbation theory in \cite{Hartle-Horowitz} and to all orders of the curvature expansion in \cite{beyond}). The extension of this range is likely to include multi-loop orders and is likely to be generalized to the (A)dS background considered below, with the state $|{IN}\rangle$ apparently coinciding with the Euclidean Bunch-Davies vacuum.

At the heuristical level the justification for this extension follows from Fig.\ref{Fig.1} depicting the compact Euclidean spacetime used as a tool for constructing the Euclidean vacuum within a well-known no-boundary prescription \cite{noboundary}. Attaching a Euclidean space hemisphere to the Lorentzian de Sitter spacetime makes it {\em compact} instead of the original asymptotic de Sitter infinity. Thus it simulates by the path integral over regular field configurations on this spacetime the effect of the Euclidean de Sitter invariant vacuum. The role of spacetime {\em compactness} is very important here because it allows one to disregards possible surface terms originating from integrations by parts or using cyclic permutations under the functional traces in the Feynman diagrammatic technique for the effective action.

In what follows this property will be very important. In particular, the Green's function will be uniquely defined by the condition of regularity on such a compact spacetime without a boundary. This information is sufficient to specify the Green's function, for which we require the following symmetric variational law (with respect to local metric variations in $\Box$ and $\hat P$)
    \begin{eqnarray}
    \delta\frac1{\Box+\hat P}=-\frac1{\Box+\hat P}\,\delta\big(\Box+\hat P\big)
    \frac1{\Box+\hat P},              \label{symvar}
    \end{eqnarray}
characteristic of the Euclidean signature d'Alembertian defined on the space of regular fields on a compact spacetime without a boundary.

A similar treatment of a nonlocal action in \cite{nonloccosm,DeffWood} was very reservedly called the "integration by parts trick" needing justification from the Schwinger-Keldysh technique. However, this trick only provides the causality of effective equations, but does not guarantee the Euclidean-Lorentzian relation (\ref{EuclidLorentz}). The latter is based, among other things, on the choice of the $|{IN}\rangle$-state.\footnote{The $f(\Box^{-1}R)$ approach to nonlocal cosmology \cite{nonloccosm,DeffWood,ParkDodelson} assumes as a first principle the existence of causal generally covariant equations of motion not necessarily derivable by variational procedure. Thus the action of \cite{nonloccosm} plays only the auxiliary role and is used merely as a tool of obtaining such equations, guaranteeing the matter stress tensor conservation. We are grateful to S. Deser and R. Woodard for clarifying this point.}

\section{The new representation of the action and stability of Einstein space backgrounds}
The action (\ref{action}) can be essentially simplified by using the compactness of Euclidean spacetime discussed above. Its new representation is based on the following local identity which is valid for an arbitrary pure trace tensor function $\varPhi_{\mu\nu}=g_{\mu\nu}\varPhi$,
    \begin{eqnarray}
    (\Box+\hat P)\,g_{\mu\nu}\varPhi=g_{\mu\nu}\left(\Box
    -\frac\alpha4\,R\right)\varPhi
    +A\,E_{\mu\nu}\varPhi.         \label{equation0}
    \end{eqnarray}
where $E_{\mu\nu}$ is a traceless part of the Ricci tensor defined by (\ref{Einsteinspace}). The nonlocal identity for the Green's function of the operator (\ref{operator}) arises if we take the scalar $\varPhi$ in the form of the nonlocal functional of another arbitrary scalar function $\varphi$
     \begin{eqnarray}
    \varPhi=\frac1{\Box-\frac\alpha4\,R}\,\varphi
    \end{eqnarray}
and act on (\ref{equation0}) by $(\Box+\hat P)^{-1}$ from the left, so that
    \begin{eqnarray}
    \frac1{\Box+\hat P}\,g_{\mu\nu}\varphi
    =g_{\mu\nu}\frac1{\Box-\frac\alpha4\,R}\,\varphi
    -\frac{A}{\Box+\hat P}\,E_{\mu\nu}
    \frac1{\Box-\frac\alpha4\,R}\,\varphi.  \label{equation}
    \end{eqnarray}

For a compact spacetime an important simplification occurs if we identify $\varphi$ with $R$ and take into account that
    \begin{eqnarray}
    \frac1{\Box-\frac\alpha4\,R}\,R
    =-\frac4\alpha .                     \label{equation1}
    \end{eqnarray}
This equation holds for a compact spacetime without a boundary or under boundary conditions which do not generate surface terms under integration by parts in the following chain of relations
    \begin{eqnarray}
    &&\frac1{\Box-\frac\alpha4\,R}\,R=-\frac4\alpha\,
    \frac1{\Box-\frac\alpha4\,R}\,
    \left(\overrightarrow{\Box}-\frac\alpha4\,R\right)1
    =-\frac4\alpha\,
    \frac1{\Box-\frac\alpha4\,R}\,
    \left(\overleftarrow{\Box}-\frac\alpha4\,R\right)1
    =-\frac4\alpha.                      \label{equation2}
    \end{eqnarray}
Therefore we have the basic identity
    \begin{eqnarray}
    \frac1{\Box+\hat P}\,g_{\mu\nu}\frac{R}4
    =-\frac1\alpha\,g_{\mu\nu}
    +\frac{A}\alpha\frac1{\Box+\hat P}\,
    E_{\mu\nu}                       \label{equation3}
    \end{eqnarray}
and two its straightforward corollaries
    \begin{eqnarray}
    &&\frac\alpha{\Box+\hat P}\,G_{\mu\nu}
    =g_{\mu\nu}
    +\frac{\alpha-A}{\Box+\hat P}\,
    E_{\mu\nu},                       \label{equation4}\\
    &&\frac\alpha{\Box+\hat P}\,R_{\mu\nu}
    =-g_{\mu\nu}
    +\frac{\alpha+A}{\Box+\hat P}\,
    E_{\mu\nu}.                       \label{equation5}
    \end{eqnarray}

Systematically using these identities in the integrand of (\ref{action}) we see that the Einstein term (linear in curvature) gets canceled and the the result becomes quadratic in $E_{\mu\nu}$
    \begin{eqnarray}
    &&S=\frac{M^2}2\int dx\,g^{1/2}\,
    \left\{-R+R^{\mu\nu}
    \left(\frac\alpha{\Box+\hat P}
    \,G_{\mu\nu}\right)\right\}\nonumber\\
    &&\qquad\qquad\qquad=
    -\frac{M^2}2\,\frac{A^2-\alpha^2}\alpha
    \int dx\,g^{1/2}\,E^{\mu\nu}
    \frac1{\Box+\hat P}\,E_{\mu\nu}.
    \end{eqnarray}
This is a new representation of the action (\ref{newrep}) which is exact and explicitly contains the effective Planck mass (\ref{effectivemass}) suggested in \cite{serendipity}.

It immediately allows one to prove the existence of a generic Einstein space solutions (including the maximally symmetric ones derived in \cite{serendipity}) and the absence of ghost modes on top of them. Since (\ref{newrep}) is quadratic in $E_{\mu\nu}$ its first order derivative is at least linear in $E_{\mu\nu}$ with some complicated nonlocal operator coefficient,
    \begin{eqnarray}
    &&\frac{\delta S}{\delta g_{\mu\nu}}=\frac{M^2_{\rm eff}}2
    \,g^{1/2}\,
    \Omega^{\mu\nu}_{\;\;\;\;\alpha\beta}(\nabla)\,
    \frac1{\Box+\hat P}\,
    E^{\alpha\beta},                         \label{eom}\\
    &&\Omega^{\mu\nu}_{\;\;\;\;\alpha\beta}(\nabla)
    =\Box\,\delta^{\mu\nu}_{\alpha\beta}
    +g^{\mu\nu}\nabla_\alpha\nabla_\beta
    -2\nabla_{(\alpha}\nabla^{(\mu}\delta^{\nu)}_{\beta)}
    +\frac12\,R\,
    \delta^{\mu\nu}_{\alpha\beta}+O[\,E\,],   \label{Omega}
    \end{eqnarray}
where $O[\,E\,]$ denotes terms vanishing in the limit $E_{\mu\nu}\to 0$. This guarantees the existence of vacuum solutions with $E_{\mu\nu}=0$. Perturbative stability of these solution follows from the quadratic part of the action, which is easily calculable now.

In view of the quadratic nature of (\ref{newrep}), the quadratic part of the action on the Einstein space background requires variation of only two explicit $E_{\mu\nu}$-factors. For the metric variations $\delta g_{\mu\nu}\equiv h_{\mu\nu}$ satisfying the DeWitt gauge
    \begin{eqnarray}
    \chi^\mu\equiv\nabla_\nu
    h^{\mu\nu}-\frac12\,\nabla^\mu h=0,  \label{DWgauge}
    \end{eqnarray}
the variation of $E_{\mu\nu}$ reads
    \begin{eqnarray}
    &&\delta E_{\mu\nu}\Big|_{\;E_{\alpha\beta}=0}=-\frac12\,\Box h_{\mu\nu}-W_{(\mu\;\;\nu)}^{\;\,(\alpha\;\;\beta)} h_{\alpha\beta}+\frac1{12}\,Rh_{\mu\nu}
    +\frac18\,g_{\mu\nu}\left(\Box-\frac16\,R\right)h =-\frac12\hat D\,
    \bar h_{\mu\nu},                     \label{deltabarR}
    \end{eqnarray}
where the operator $\hat D$
    \begin{eqnarray}
    \hat D\equiv \Box+2\hat W
    -\frac16\,R\,\hat1,       \label{D}
    \end{eqnarray}
acts on a traceless part of $h_{\mu\nu}$, the hat labels matrices acting on symmetric tensors,
    \begin{eqnarray}
    &&\bar h_{\mu\nu}\equiv
    \hat\varPi h_{\mu\nu}=h_{\mu\nu}
    -\frac14\,g_{\mu\nu}h,\quad
    \hat\varPi\equiv
    \varPi_{\mu\nu}^{\;\;\alpha\beta}
    =\delta_{\mu\nu}^{\alpha\beta}
    -\frac14\,g_{\mu\nu}g^{\alpha\beta},\\
    &&\hat W h_{\mu\nu}\equiv W_{(\mu\;\;\nu)}^{\;\,(\alpha\;\;\,\beta)}
    h_{\alpha\beta},
    \end{eqnarray}
and $W_{\mu\;\nu}^{\;\alpha\;\beta}$ denotes the Weyl tensor.
Note that the operator $\hat D$ commutes with the projector $\hat\varPi$, $[\hat\varPi,\hat D]=0$, because of the traceless nature of the Weyl tensor, $\hat\varPi\hat W=\hat W\hat\varPi=\hat W$, so that the variation (\ref{deltabarR}) of the traceless $E_{\mu\nu}$ is also traceless as it should.

In matrix notations the operator $\Box+\hat P$ on the Einstein background reads
    \begin{eqnarray}
    \big(\Box+\hat P\big)\Big|_{\;E_{\mu\nu}=0}=\Box+a\,\hat W-\frac{C}4\,R\hat\varPi
    -\frac\alpha4\,R\,(\hat1-\hat\varPi).
    \end{eqnarray}
Therefore, in view of (\ref{deltabarR}), the property $[\hat\varPi,\hat D]=0$ and the obvious relation
    \begin{eqnarray}
    \hat\varPi\,\frac1{\Box+\hat P}\,\hat\varPi=\hat\varPi\,\frac1{\Box+a\,\hat W-\frac{C}4\,R\,\hat1}\,\hat\varPi
    \end{eqnarray}
we finally have the quadratic part of the action in terms of the traceless part $\bar h_{\mu\nu}$ of the metric perturbations $h_{\mu\nu}$ satisfying the DeWitt gauge
    \begin{eqnarray}
    S_{(2)}\Big|_{\;E_{\mu\nu}=0}
    =-\frac{M^2_{\rm eff}}2
    \int d^4x\,g^{1/2}\big(\hat D\bar h^{\mu\nu}\big)
    \,\frac1{\Box+a\,\hat W
    -\frac{C}4\,R\,\hat1}\,
    \big(\hat D\bar h_{\mu\nu}\big).  \label{S_2}
    \end{eqnarray}
This expression was first derived in \cite{Solodukhin}.

For generic values of the parameters $a$ and $C$ the propagator of the theory features double poles corresponding to the zero modes of the operator $\hat D$. This is a nonlocal generalization of the situation characteristic of the critical gravity theories with a local action containing higher-order derivatives \cite{critical}. Local theories with double poles have a distinguished status different from unstable higher-derivative models with massive ghosts -- their stability is determined also by special logarithmic modes which might or might not violate unitarity \cite{critical}. Interestingly, flexibility in the values of the parameters $a$ and $C$ allows us to avoid perturbative instability of the Einstein space background. The quadratic form (\ref{S_2}) can be made local and thus guarantee the existence of the propagator with a single positive-residue pole. This is easily achieved by demanding equality of the operator (\ref{D}) and the operator in the denominator of (\ref{S_2}) along with the positivity of $M^2_{\rm eff}$,
    \begin{eqnarray}
    \hat D=\Box+a\,\hat W-\frac{C}4\,R\,\hat1.
    \end{eqnarray}
This yields the value $C=2/3$ derived in \cite{serendipity} by very extensive calculations and in addition leads to a unique value of another parameter $a=2$, which allows us to extend stability arguments to generic Einstein space backgrounds \cite{Solodukhin} (the condition $a=2$ is not necessary on maximally symmetric background with $\hat W=0$ and, thus, was derived in \cite{Solodukhin} in the course of generalizing the model of \cite{serendipity} to generic Einstein spaces).

\section{GR phase: asymptotically flat spacetime vs cosmological boundary conditions}
Using (\ref{Omega}) in the equation of motion (\ref{eom}) one can see that in the UV limit $\nabla\nabla\gg R$ the variational derivative of the action
    \begin{eqnarray}
    &&\frac{\delta S}{\delta g_{\mu\nu}}\simeq
    \frac{M^2_{\rm eff}}2
    \,g^{1/2}\left(R_{\mu\nu}
    -\frac12\,\nabla_\mu\nabla_\nu\frac1\Box R\right)
    +O[\,E^2\,]
    \end{eqnarray}
remains nonlocal and differs from the general relativistic expression even for $\alpha\to 0$. In particular, in the approximation linear in the curvatures matter sources are coupled to gravity according to
    \begin{eqnarray}
    R_{\mu\nu}
    -\frac12\,\nabla_\mu\nabla_\nu\frac1\Box R
    +O[\,R^2\,]
    =\frac1{M^2_{\rm eff}}\,T_{\mu\nu}, \label{mattersource}
    \end{eqnarray}
where nonlinear in the curvature terms $O[\,R^2\,]$ include nonlinearity in $E_{\mu\nu}$. The local Ricci scalar term of the Einstein tensor is replaced here with the nonlocal expression which guarantees in this approximation the stress tensor conservation, but in contrast to anticipations of \cite{serendipity} does not provide the GR phase of the theory.

The absence of the GR phase might seem paradoxical because the original action (\ref{action}) obviously reduces to the Einstein one in the limit $\alpha\to 0$. The explanation of this paradox consists in the observation that the transition from (\ref{action}) to the new representation (\ref{newrep}) is based on the identity (\ref{equation1}) which is not analytic both in $\alpha$ and in the curvature. The source of this property is the constant zero mode of the scalar operator $\Box$ on compact Euclidean spacetimes without a boundary. On such manifolds the left hand side of (\ref{equation1}) is not well defined for $\alpha=0$. The equivalence of the actions (\ref{action}) and (\ref{newrep}) was obtained only on this class of Euclidean manifolds. The latter, in turn, were motivated in Sect.2 by extending the duality between the Schwinger-Keldysh technique and Euclidean QFT \cite{beyond} to the cosmological (quasi-de Sitter) context.

In contrast to this class of manifolds, the representations (\ref{action}) and (\ref{newrep}) are not equivalent in asymptotically flat (AF) spacetime because equations (\ref{equation1})-(\ref{equation5}) do not apply there. First, with zero boundary conditions at infinity the scalar $\Box$ does not have zero modes. Second, due to the natural AF falloff conditions, $R(x)\sim 1/|x|^4$ and $(1/\Box)\delta(x-y)\sim 1/|x-y|^2$, integration by parts in the chain of transformations (\ref{equation2}) gives a finite surface term at infinity $|x-y|\to\infty$. This leads to an alternative equation
    \begin{eqnarray}
    \frac1{\Box-\frac\alpha4\,R}\,R\,\Big|_{\,\rm AF}
    =O\,[\,R\,]                      \label{equation11}
    \end{eqnarray}
with a nontrivial right hand side analytic in $\alpha$ and tending to zero for a vanishing scalar curvature. This explains why the model (\ref{action}) on AF background has a good GR limit with nonlinear curvature corrections controlled by a small $\alpha$ \cite{covnonloc,serendipity}.\footnote{Basic example of a physically nontrivial Einstein space is the Schwarzchild-de Sitter background. A priori it can also generate surface terms in (\ref{equation2}), because its metric is not smooth simultaneously at the black hole and cosmological horizons and has a conical singularity \cite{GibHawkPage}. However, one can show that for any type of boundary conditions at this singularity the relevant surface term vanishes and leaves Eq.(\ref{equation1}) intact. A similar issue remains open in the case of the Schwarzchild-AdS background for which the operator $\hat D$ with $R<0$ is not guaranteed to be free of zero modes and does not provide a well defined representation (\ref{newrep}) \cite{Solodukhin}. We are grateful to S. Solodukhin for a discussion of this point.}

This undermines the utility of the model (\ref{action}) as a possible solution of the dark energy problem and simulation of dark matter phenomenon advocated in \cite{serendipity}. Absence of the GR limit for $\alpha\to 0$ and for short distance regime $\nabla\nabla\gg R$ becomes a critical drawback of this model\footnote{In \cite{Solodukhin} this was interpreted as the phase transition between the $R=4\Lambda>0$ and $R=0$ phases -- the absence of crossover between these phases. We see that in fact this transition has a topological nature.} caused by its infrared behavior -- presence of a constant zero mode on a compact spacetime. Possible solution of this problem could be a reformulation of the nonlocal action by projecting out this zero mode from the definition of the Green's function in (\ref{action}) (see \cite{Zelnikovetal} for the technique of such a truncation).

Another possible way to circumvent this difficulty can be based on the conformal transformation to a new metric
    \begin{eqnarray}
    \tilde g_{\mu\nu}[\,g\,]=e^{2\sigma[\,g\,]}\,g_{\mu\nu},
    \end{eqnarray}
which is assumed to be physical (that is directly coupled to matter) in contrast to the original metric $g_{\mu\nu}$ playing the auxiliary role. With the conformal factor function
    \begin{eqnarray}
    \sigma[\,g\,]\simeq\frac14\,\frac1\Box R,
    \end{eqnarray}
which is small in the UV limit, $\sigma\ll 1$, but has large second order derivatives\footnote{Note that this expression is assumed to hold only in the formal UV limit of $\nabla\nabla\gg R$, so that the zero mode of $\Box$ should not invalidate it.}, $\nabla\nabla\sigma\sim R$, one can express the covariant Einstein tensor of the new metric $\tilde G_{\mu\nu}$ in terms of the original metric as
    \begin{eqnarray}
    &&\tilde G_{\mu\nu}=G_{\mu\nu}+2\big(g_{\mu\nu}\Box\sigma
    -\nabla_\mu\nabla_\nu\sigma\big)+g_{\mu\nu}\sigma_\alpha^2
    +2\sigma_\mu\sigma_\nu\nonumber\\
    &&\qquad\qquad=R_{\mu\nu}
    -\frac12\,\nabla_\mu\nabla_\nu\frac1\Box R+O\left[\,\Big(\nabla\frac1\Box R\Big)^2\right], 
    \quad \sigma_\mu\equiv\nabla_\mu\sigma.
    \end{eqnarray}
We see that $\tilde G_{\mu\nu}$ in this limit in fact reproduces the left hand side of (\ref{mattersource}). Therefore, if we couple matter to the new metric $\tilde g_{\mu\nu}$ in the total action as
    \begin{eqnarray}
    S_{\rm total}[\,g,\phi\,]=S[\,g\,]+S_{\rm matter}[\phi,\tilde g[\,g\,]\,],
    \end{eqnarray}
then for $\tilde g_{\mu\nu}$ in the short distance limit we will recover the usual Einstein equations
    \begin{eqnarray}
    \tilde R_{\mu\nu}
    -\frac12\,\tilde g_{\mu\nu}\tilde R
    =\frac1{M^2_{\rm eff}}\,\tilde T_{\mu\nu},
    \quad
    \tilde T_{\mu\nu}=\frac2{\tilde g^{1/2}}\,\tilde g_{\mu\alpha}\tilde g_{\nu\beta}\,\frac{\delta S_{\rm matter}}{\delta\tilde g_{\alpha\beta}}
     \label{modeq}
    \end{eqnarray}
where $\tilde T_{\mu\nu}$ is a matter stress tensor in the frame of the $\tilde g_{\mu\nu}$-metric. When deriving this equation we took into account smallness of $\sigma$ and $\delta\sigma/\delta g_{\mu\nu}=O(\sigma)$ in the short distance limit $\nabla\nabla\gg R$. Thus we get a GR phase in the conformally related frame of the theory. Unfortunately, however, the magnitude of corrections to the GR behavior is no longer controlled by a small parameter $\alpha$, which makes application of this idea to realistic cosmology still somewhat questionable.

\section{Conclusions}
We have derived the equivalent representation (\ref{newrep}) of the action (\ref{action}) with the critical value (\ref{relation}) of the parameter $\alpha$. This representation allows one in a systematic way to extend applications of these models from maximally symmetric to generic Einstein spaces and black hole solutions. Unfortunately, in contrast to AF spacetimes this model fails to have a general relativistic limit in the cosmological problems for the mean metric field, treated within the Euclidean version of the Schwinger-Keldysh formalism. Nevertheless, the short-distance GR limit can be attained in a special conformal frame (physical metric minimally coupled to matter) nonlocally related to the original one. This limit, however, cannot be controlled by smallness of the parameter $\alpha$ that was initially designed in \cite{serendipity} to moderate the effect of nonlocal corrections to the Einstein theory.

Thus, direct cosmological implications of the model (\ref{action}) are not likely to be available. However, it might be interesting as a nonlocal generalization of critical gravity theories \cite{critical} which recently became popular as holographic duals of the logarithmic conformal models \cite{GrumillerSachs}. In fact, the relation (\ref{relation}) can be regarded as the analogue of the criticality condition in the local quadratic in curvature models. It eliminates massive gravitons and gives rise to logarithmic modes \cite{critical} corresponding to the double pole in the propagator. Zero energy of massless gravitons and positive energy of log modes \cite{critical} give rise to controversial statements on unitarity of these critical models (see the work \cite{PorratiRoberts} claiming the loss of unitarity due to lack of orthogonality between the logarithmic and Einstein states). Analogous reasoning might imply that our model is also stable even without imposing the conditions (\ref{Crelation}) and (\ref{arelation}). In fact, the theory (\ref{newrep}) bears a number of properties in common with critical gravity models of \cite{critical}. In particular, as advocated in \cite{Solodukhin}, it has Schwarzschild-de Sitter black hole solutions with zero entropy in parallel with zero entropy and energy black holes of \cite{critical}. All this makes the class of nonlocal gravity models open for interesting future implications.

\section*{Acknowledgements}
The authors strongly benefitted from fruitful discussions and correspondence with S. Deser, S. Solodukhin, R. Woodard and A. Zelnikov.  This work of A. B. was supported by the RFBR grant No. 11-01-00830. The work of Yu. G. was supported by the RFBR grant No 11-02-00512 and the grant from FAPEMIG.

\end{document}